\documentclass[10pt,aps,prx,twocolumn,notitlepage,nofootinbib,superscriptaddress]{revtex4-1}
\usepackage{amsmath,amsfonts,amssymb}
\usepackage{graphicx}

\begin{document} 
\title{Interevent time distribution, burst, and hybrid percolation transition}

\author{Jinha Park}
\affiliation{CCSS, CTP and  Department of Physics and Astronomy, Seoul National University, Seoul 08826, Korea}
\author{Sudo Yi}
\affiliation{CCSS, CTP and  Department of Physics and Astronomy, Seoul National University, Seoul 08826, Korea}
\affiliation{School of Physics, Korea Institute for Advanced Study, Seoul 02455, Korea}
\author{K. Choi}
\author{Deokjae Lee}
\author{B. Kahng}
\email{bkahng@snu.ac.kr}
\affiliation{CCSS, CTP and  Department of Physics and Astronomy, Seoul National University, Seoul 08826, Korea}

\date{\today}

\begin{abstract}
Critical phenomena of a second-order percolation transition are known to be independent of cluster merging or pruning process. However, those of a hybrid percolation transition (HPT), mixed properties of both first-order and second-order transitions, depend on the processes. The HPT induced by cluster merging is more intrigue and little understood than the other. Here, we construct a theoretical framework using the so-called restricted percolation model. In this model, clusters are ranked by size and partitioned into small- and large-cluster sets. As the cluster rankings are updated by cluster coalescence, clusters may move back and forth across the set boundary. The inter-event time (IET) between two consecutive crossing times have two distributions with power-law decays, which in turn characterize the criticality of the HPT. A burst of such crossing events occurs and signals the upcoming transition. We discuss a related phenomenon to this critical dynamics.
\end{abstract}

\maketitle 
\section{Introduction}
Complex systems composed of a large number of interacting components are constantly adapting to the changing environment and sometimes reach critical states~\cite{complex1,complex2}. In these states, complex systems may be statistically characterized by power-law distributions~\cite{sornette2012}. Examples include the number of aftershocks per time step after the main earthquake~\cite{olami}, nonstationary relaxation after a financial crash~\cite{finance}, the number of fires with a certain area per time step as a function of the area in forest fires~\cite{forest,djlee}, and the interevent time (IET) distribution in human activities~\cite{burst}. Such critical phenomena occur in a self-organized manner, but a single framework describing their underlying mechanisms has not yet been established ~\cite{sornette2012,prediction,prediction_sornette}. In contrast, in critical states of complex systems, a small external perturbation imposed on a system can lead to widespread failure of the system through avalanche dynamics~\cite{parisi}. Examples include blackouts in electric power-grid systems~\cite{blackout,dsouza,motter} and firing in neuronal networks~\cite{cognition,herrmann} in real-world systems, and $k$-core percolation~\cite{kcore,kcore_lee,kcore_mendes} and disease contagion models~\cite{contagion1,contagion2,golden} in artificial model systems. In such cascade dynamics, the avalanche sizes of different events form a power-law distribution. These behaviors have been explained by a universal mechanism (the critical branching process), which was also observed in self-organized criticality in the Bak, Tang, and Wiesenfeld (BTW) model~\cite{btw,btw_sf}.

Avalanche dynamics can be clearly observed in $k$-core percolation. The $k$ core of a network is a subgraph, in which degree of each node is at least $k$. To obtain a $k$-core subgraph, once an Erd\H{o}s--R\'enyi (ER) random network composed of $N$ nodes having links between two nodes with probability $p$ is generated in the supercritical regime, all nodes with degrees less than $k$ are deleted consecutively until no more nodes with degrees less than $k$ remain in the system. The number of nodes deleted during these consecutive pruning processes is considered as the avalanche size. The fraction of nodes remaining in the largest $k$-core subgraph is defined as the order parameter $m$, which decreases continuously with decreasing $p$ at criticality. When $p$ is chosen as a transition point $p_c$, the deletion of a node from an ER network can lead to collapse of the giant $k$-core subgraph. Thus, a hybrid percolation transition (HPT) occurs, which exhibits features of first-order and second-order phase transitions. The avalanche size distribution shows power-law decay as $P_a(s)\sim s^{-\tau_a}$, where $\tau_a=3/2$ at $p_c$, similar to that in the BTW model. Furthermore, there exists a critical avalanche of size $O(N)$, which may correspond to the dragon king often noted in complex systems~\cite{sornette2012,souza_dk}.

In contrast to the HPTs induced by cascading dynamics, HPTs in cluster merging dynamics (CMD) have received little attention. In this paper, we aim to investigate the underlying mechanism of the HPT on a microscopic scale using the so-called restricted Erd\H{o}s--R\'enyi ($r$-ER) network~\cite{r-er,cho}. Then, we set up a theory of the HPT. The model contains a factor that suppresses the growth of large clusters. Accordingly, the type of percolation transition is changed to a discontinuous transition~\cite{riordan}, similar to the case in which the $1/r^2$-type long-range interaction changes the transition type to the first-order type in the one-dimensional Ising model~\cite{thouless}. However, this factor does not guarantee the occurrence of critical behavior. Here, we uncover a underlying mechanism and find another factor that governs the critical behavior of the HPT induced by the CMD.

This paper is organized as follows: In Sec.~\ref{sec:rer}, we recall a $r$-ER model for the HPT.  In Sec.~\ref{sec:ce2hpt}, we exmaine the cluster evolution of the $r$-ER model macroscopically and microscopically. Then we characterize the size distribution of finite clusters by constucting a theoretical framework analogous to that of the conventional percolation theory. In Sec.~\ref{sec:iet}, we introduce inter-event times and their distribution, and then show that the inter-event time distribution exhibit two power-law decays depending on time windows. We obtain the two exponents of the IET distribution analytically and numerically and then show that these exponents are related to the exponents characterizing the HPT. In Sec.~\ref{sec:discussion}, we introduce unconventional patterns that emerge together with the HPT such as a burst and the Devil's staircase and discuss an application of the model. A summary is presented in Sec.~\ref{sec:conclusion}.

\section{$r$-ER Model}
\label{sec:rer}

The CMD of the $r$-ER model proceeds in a dichotomous and asymmetric way. Initially, there are $N$ isolated nodes. Clusters are formed as links are connected one-by-one between pairs of unconnected nodes under the rule given below. Clusters are ranked by size and classified into two sets, $A$ and $B$, which contain a portion $gN$ ($0 < g \le 1$) of nodes of the smallest clusters and the remaining large clusters, respectively. Two nodes are selected for connection as follows. One node is chosen randomly from set $A$, and the other is chosen from among all the nodes. They are connected by a link unless they are already connected. Then the classification is updated as the cluster rankings are changed. Time is defined as $t=L/N$, where $L$ is the number of occupied links, and it is the control parameter of the $r$-ER model. This quantity differs from $p$ in $k$-core percolation in the point that the CMD at a certain time $t_1$ successively follows the dynamics of  previous times, whereas the avalanche dynamics at a certain $p_1$ may be independent of those at any  previous $p > p_1$.

The nodes in set $A$ have twice the opportunity to be linked compared to the nodes in set $B$; small clusters in set $A$ grow rapidly, and the resulting large clusters may move to $B$, whereas the smallest clusters among the clusters in set $B$ that have never grown or are growing slowly are evicted to $A$. Accordingly, cluster coalescence occurs in a dichotomous way, and the growth of large clusters is practically suppressed. The effective suppression becomes global as the portion of the smallest clusters is selected from among clusters of all sizes. 
This factor leads to a discontinuous percolation transition.

\begin{figure*}[!th]
\includegraphics[width=\linewidth]{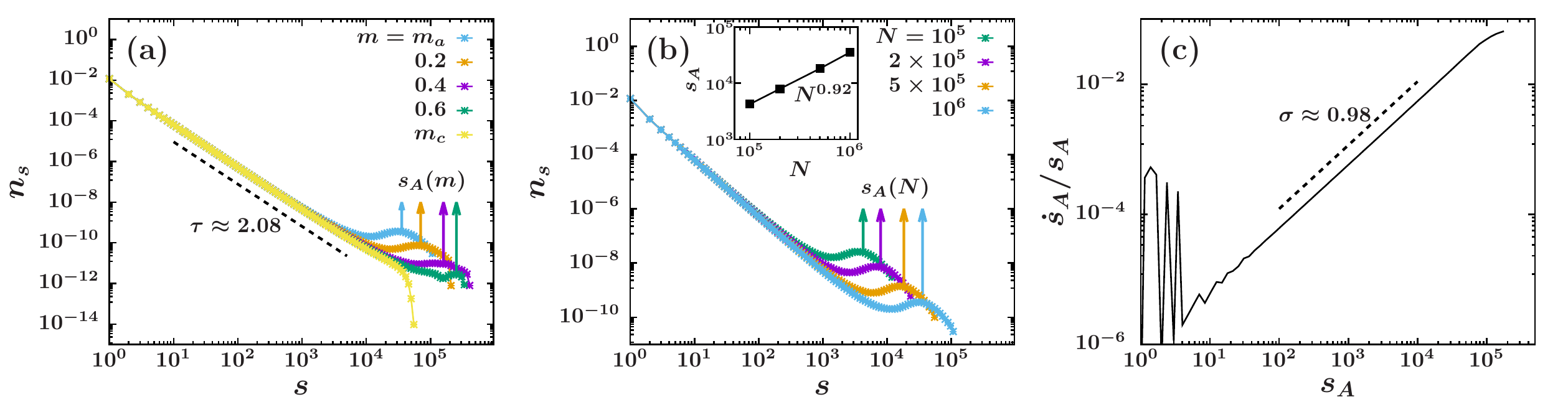}
\caption{\textbf{Characterization of cluster size distribution:} (a) Cluster size distributions $n_s$, which is controlled by the corresponding cluster size cutoff $m$, at various times. $s_A$ is located around the peak position of the bump for each time. $g=0.2$ and $N=10^6$. (b) System size dependence of $n_s$ at $m_a$. The mean $s_A$ grows with the system size $N$. Inset: $s_A(m_a)$ depends on $N$ as $\sim N^{0.92}$. (c) To determine the exponent $\sigma^\prime$ of $s_A(t)\sim (t_g-t)^{-1/\sigma^\prime}$, we plot ${\dot s}_A/s_A$ versus $s_A$. $\sigma^\prime\approx 0.98$ is obtained. Ensemble average is taken over $10^5$ configurations.}
\label{fig:tg}
\end{figure*}
\begin{figure*}
\includegraphics[width=\linewidth]{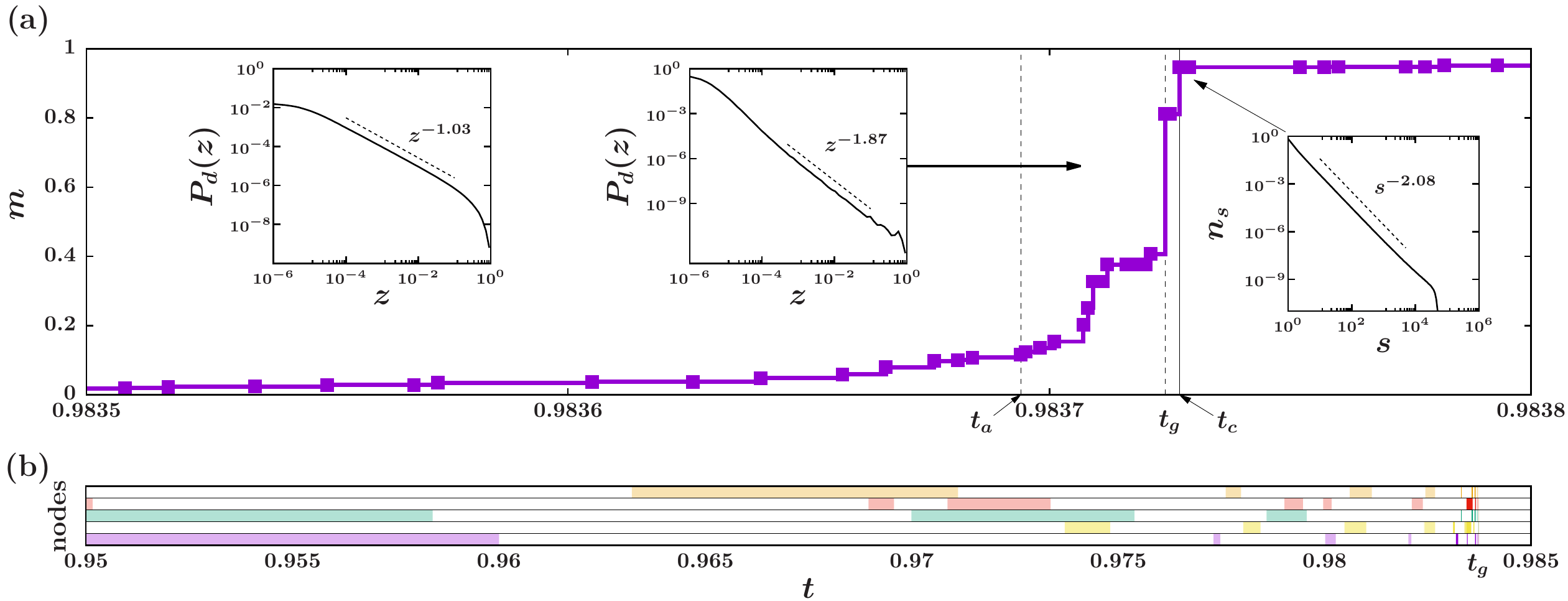}	
\caption{\textbf{Evolution of the giant cluster, IET, and burst:} (a) Typical staircase growth of the largest cluster size per node $m(t)$ for $g=0.2$. The giant cluster of size $O(N)$ forms at time $t_a$ and grows rapidly. Soon afterwards, at $t_g$, it completely fills set $B$. Subsequently, the system follows ER dynamics and reaches a transition point $t_c$, at which $n_s(t_c)\sim s^{-\tau}$ (right inset) for finite clusters. The IET distributions $P_d(z)$ are accumulated in the intervals $[0,t_a]$ (left) and $[t_a,t_g]$ (middle), respectively. They exhibit power-law decays with the exponents $\alpha\approx 1.03$ and $1.87$, respectively. (b) A succession of inter-set $A$ (colored) $\leftrightarrow$ $B$ (white) switching events demonstrate the bursty nature of this heavy-tailed process.} 
\label{fig:event}
\end{figure*} 

\section{Cluster evolution to HPT}
\label{sec:ce2hpt}
We first consider cluster evolution on a macroscopic scale. In the early time regime, the order parameter $m(t)$ (the fraction of nodes belonging to the giant cluster) is $o(N)$, and the cluster size distribution $n_s(t)$ exhibits power-law decay in the small-cluster region, but exponential decay in the large-cluster region. As time proceeds, medium-size clusters accumulate and form a bump in the cluster size distribution $n_s(t)$, as shown in Fig.~\ref{fig:tg}(a) and Appendix~\ref{app:n_s}. Technically, we trace $m$ instead of $t$ in simulations to reduce large sample fluctuations arising around the transition point of the HPT. We estimate a characteristic time $t_a$, around which a giant cluster of size $m_aN\sim O(N)$ emerges and the bump size becomes maximum. $m_a$ denotes $m(t_a)$. Soon after that, $m(t)$ increases rapidly, as shown in Fig.~\ref{fig:event}(a). $t_a$ is estimated numerically as explained in Appendix~\ref{app:m_a}. At $t_g$, the order parameter $m(t_g)$ (denoted as $m_g$) is equal to $1-g$. This means that set $B$ is occupied by the giant cluster alone. Beyond $t_g$, the giant cluster size exceeds the capacity of set $B$. In this case, the giant cluster is regarded as belonging to set $A$, along with all the other clusters~\cite{cho}. The boundary between the two sets is eliminated. Note that this rule was not clearly specified in the original half-restricted model~\cite{r-er}; however, this rule is necessary to reach a critical state at $t_c$.  Therefore, the $r$-ER model reduces to the original ER model but with the $n_s(t_g)$ of finite clusters and the giant cluster~\cite{cho}. As time proceeds further to a transition point $t_c$, the bump disappears completely, and the size distribution of finite clusters $n_s$ exhibits power-law decay, $n_s(t_c)\sim s^{-\tau}$ (Fig.~\ref{fig:event}(a)), where $\tau(g)$ varies continuously with the parameter $g$~\cite{cho}. The order parameter at $t_c$ is denoted as $m_c$. The interval $[t_a, t_c]$ has been revealed to be $o(1)$~\cite{r-er}. Therefore, the order parameter is regarded as discontinuous at $t_c$ in the limit $N\to \infty$. For $t > t_c$, the order parameter increases continuously with the critical behavior, $m(t)-m_c\sim (t-t_c)^{\beta}$. A hybrid transition occurs in the order parameter. We divide the time interval $[0,t_g]$ into two windows. In $[0,t_a]$, $n_s$ exhibits power-law decay with an extra bump, whereas during $[t_a,t_g]$, the bump shrinks, and the giant cluster grows dramatically. These behaviors clearly indicate that these two intervals need to be considered separately. 

Next, we consider cluster evolution microscopically. We check the evolution of the largest cluster size among all the clusters in set $A$, denoted as $s_A(t)$. We find empirically that $s_A \sim (t_g-t)^{-1/\sigma^\prime}$ for $t\in [0,t_a]$, which is analogous to the relationship used in conventional percolation theory. We measure the exponent $\sigma^\prime$ directly in Appendix~\ref{app:n_s} and using the relation ${\dot s}_A/s_A \sim s_A^{\sigma^\prime}$ (Fig.~\ref{fig:tg}(c)), where the dot over $s_A$ represents the time derivative. The exponent $\sigma^\prime(g)$ is almost one for $g=0.2$ but decreases slowly with increasing $g$, as shown in Appendix~\ref{app:n_s}. Note that the value of $\sigma^\prime$ differs from $\sigma$ that is  conventionally defined  for the characteristic cluster size in the supercritical regime, where $\sigma=1$ independent of $g$~\cite{cho}. Interestingly, $s_A(t)$ is located around the peak position of the bump in $n_s(t)$ for different $t$ but is implemented using the corresponding $m$ value, as shown in Fig.~\ref{fig:tg}(a). This behavior appears for different system sizes, as shown in Fig.~\ref{fig:tg}{(b)}. The peak at $s_A$ implies that clusters with sizes similar to $s_A(t)$ are abundant in the system even though the bump shrinks as $t$ increases beyond $t_a$.  

\begin{figure}
\centering\includegraphics[width=0.8\linewidth]{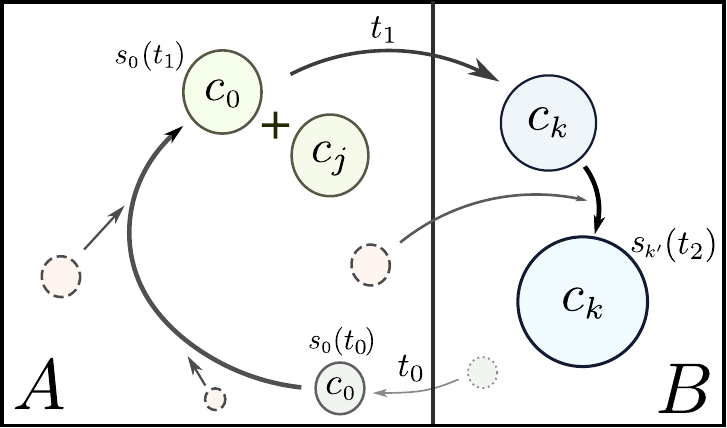} 
\caption{\textbf{Schematic diagram of a typical cluster lifecycle:} A cluster is born with size $s_0$ and grows through coalescence. Each constituent node of a cluster may have a different age (time since its birth). After sufficient growth in set $A$, a cluster is transferred to set $B$, where its growth is limited.}
\label{fig:schematic}
\end{figure}
More specifically, we consider how a cluster $c_i$ of size $s_i$ evolves during the bump formation period $[0,t_a]$. A schematic illustration is presented in Fig.~\ref{fig:schematic}.   

\begin{enumerate} 
\item[i)] Suppose that a cluster $c_0$ is evicted from set $B$ to $A$ at time $t_0$. If $t_0=1$, cluster $c_0$ has size $s_0=1$ in set $A$. Next, $c_0$ is merged with other clusters and grows, but it is still small enough that it does not move to $B$. 
\item[ii)] Until time $t_1$, cluster $c_0$ grows and has size $s_0(t_1)$. At time $t_1$, cluster coalescence occurs as $c_0+c_j\to c_k$, and the cluster size changes as $s_k=s_0(t_1)+s_j$, where cluster $c_j$ is in either set $A$ or set $B$. We consider the case that the cluster size $s_k$ becomes larger than $s_A(t_1)$, and thus cluster $c_k$ moves to $B$. 
\item[iii)] Cluster $c_k$ grows slowly to the size $s_{k'}$ in set $B$ until time $t_2$. If the size $s_{k'}$ becomes smaller than $s_A(t_2)$ for the first time, then cluster $c_k$ is evicted to $A$. Then the evolution returns to step i). 
\end{enumerate}
This cycle is a prototypical pattern of cluster evolution up to the time $t_a$. During this cycle, small clusters (e.g., $c_0$) in set $A$ have more opportunities to grow, whereas the growth of large clusters in set $B$ (e.g., $c_k$) is suppressed. Accordingly, clusters of medium size become abundant and form a bump around $s_A(t)$, as shown in Fig.~\ref{fig:tg}(a).  

During the interval $t_a < t < t_g$, a non-negligible amount of cluster coalescence occurs between a large cluster in set $A$ and another large cluster in set $A$ or the giant cluster in set $B$. In this case, step iii) must also include the following event.    

\begin{enumerate}
\item[iii$'$)] Cluster $c_k$ of size $s_{k'}$ grows further by merging with clusters in set $A$, and $s_{k'}(t) \le s_A(t)$ never occurs through $t_g$.  
\end{enumerate}

\section{Inter-event time distribution}
\label{sec:iet}

The cycle of steps i)--iii) is analyzed in terms of the duration time~\cite{burst}, which indicates how long a node remains in one set before switching to the other. Whenever a node $i$ switches from one set to the other, the duration time is reset to zero. The update process of the duration time is described in detail in Appendix~\ref{app:iet}. In Fig.~\ref{fig:event}(b), the horizontal axis denotes time, and the boundary between two domains (indicated by alternating use of color) represents an event in which a cluster of a given node $i$ moves from one set to the other set. The interval between two consecutive boundaries is called the duration time (or IET of node $i$) and is denoted as $z_i$. For each node $i$, there can exist many duration times generated by those set-crossing events. The IET distribution $P_d(z)$ can be constructed by accumulating these duration times over all nodes during a given time interval, for instance, $[0,t_g]$. We find that the IET distribution exhibits power-law decay as $P_d(z)\sim z^{-\alpha}$ (inset of Fig.~\ref{fig:event}(a)). The exponent $\alpha$ depends on the time window. For $[0, t_a]$, $\alpha$ is measured to be approximately one for $g=0.2$, being insensitive to $g$ as long as $g$ is not close to one. This exponent value is also insensitive to the set-crossing type (either from $A$ to $B$ or from $B$ to $A$). During the window $[t_a, t_g]$, the exponent $\alpha$, alternatively denoted as $\alpha'$, is found to be $\alpha^{\prime}=4-\tau$ for the type $A \to B$, and is less than two. However, for the type $B \to A$, we obtain $\alpha' > 2$. We note that in the latter case, the duration times that were reset to zero before $t_a$ remain, even though they are measured during the window $[t_a, t_g]$. Because the IET distribution for the window $[t_a, t_g]$ decays rapidly, the exponent $\alpha$ measured during the entire period $[0,t_g]$ is governed by the values measured during the window $[0,t_a]$, and thus, the exponent $\alpha$ is close to one. 

We obtain the IET distribution analytically. Let us consider the case in which a cluster $c_0$ of size $s_0$ is evicted from $B$ to $A$ at time $t_0 > 0$. A node belonging to cluster $c_0$ has more opportunity to be selected. Thus, cluster $c_0$ grows rapidly in set $A$, and it returns to $B$ for the first time at a time $t_1=t_0+z$, at which its size becomes larger than $s_A(t_1)$. Then all $s_0$ nodes that belonged to the original cluster $c_0$ at time $t_0$ have duration time $z$, which is accumulated in the IET distribution. The probability $P_{d}^{A\to B}(z)$ that such events happen during the interval $[0,t_g]$ is calculated as  
\begin{widetext}
\begin{equation}
P_d^{A\to B}(z) = \int_{0}^{t_g} dt_0~s_0 q_1(s_0;t_0)\delta\big(s_0-s_A(t_0)\big)\bigg[\prod_{t=1/N}^{z-1/N} q(s_{t_0+t};t_0+t)\bigg]\big[1-q(s_{t_0+z};t_0+z)\big], \label{eq:pd}
\end{equation}
\end{widetext}
where $q_1(s_0;t_0)$ is the probability that a cluster of size $s_0$ in set $B$ is evicted to set $A$ at $t_0$ by an event in which a cluster larger than $s_0$ moves from $A$ to $B$ at $t_0$. 
Further, $q(s_{t_0+t};t_0+t)$ is the probability that cluster $c_0$ remains in set $A$ with size $s_{t_0+t}$ at time $t_0+t$. 
Next, we use the relation ${s_A}/{{\dot s}_A}=\sigma^\prime {s_A^{-\sigma^\prime}}$ and regard $n_s$ as $\sim s^{-\tau}/{(1-s_A^{2-\tau})}$ for $s \le s_A$. Then we obtain $P_{d}(z)\sim z^{-(4-\tau-\sigma^\prime)}$.
A detailed derivation of the above equation is presented in Appendix~\ref{app:iet}. 

We also obtain the above result using simple dimensional analysis~\cite{btw}. We calculate the average waiting time of a cluster to be selected randomly for cluster merging, which is found to be inversely proportional to the cluster size $s$, as explained in Appendix. Thus, the expected duration time of a node belonging to a cluster of size $s$ is $\langle z_s \rangle \sim 1/s$. The IET distribution is related to the cluster size distribution $n_s$ as  
\begin{equation}
P_d(z) \sim s \cdot sn_s \left|\partial s/\partial \langle z \rangle \right|,
\label{eq:3}
\end{equation}
where the factor $s$ is also needed because the IET is counted per node. The factor $sn_s$ is the probability of selecting a node in a cluster of size $s$.    
During the entire period $[0,t_g]$, the mean cluster density $\langle n_s(t)\rangle \approx (1/t_g)\int_0^{t_g}dt~s^{-\tau}f({s/s^*(t)}) \propto s^{-\tau-\sigma^\prime}$. Here $s^*(t)\sim s_A(t) \sim (t_g-t)^{-1/\sigma^\prime}h((t_g-t_a)/(t_g-t))$, where $h(x)$ is a scaling function defined as $h(x)\sim s^{-1/\sigma^\prime}$ for $x < 1$ and may be regarded as constant for $x > 1$. Therefore, $P_d(z) \sim z^{-\alpha}$, with $\alpha=4-\tau-\sigma^\prime$. Next, we recall the previous result, $\chi_R\equiv (1/g)\sum_{s=1}^{s_A}s^2n_s(t)\sim (t_c-t)^{-1}.$ This result was obtained using the Smoluchowski equation for cluster evolution~\cite{cho}. Assuming that $n_s(t)$ decays in a pure power-law manner as before, we obtain the relation $\chi_R\sim (t_c-t)^{-(3-\tau)/\sigma^\prime}$, which leads to the scaling relation $\tau+\sigma^\prime=3$, and the exponent $\alpha$ becomes $\alpha=1$. Interestingly, even though the exponents $\tau$ and $\sigma^\prime$ depend on the control parameter $g$, the exponent $\alpha=1$, independent of $g$ and universally. We note that the susceptibility exponent, $\gamma=(3-\tau)/\sigma^\prime=1$, is obtained from set $A$. 
However, we remark that the limiting case $g\rightarrow 1$ of the restricted model reduces to an ordinary ER model, in which $\tau=5/2$ and $\sigma^\prime=1/2$. Thus, $\tau+\sigma^\prime=3$ holds. For this case, we find that $\langle n_s(t)\rangle_t\equiv\frac{1}{t_c}\int_0^{t_c}n_s(t)dt\sim s^{-\tau-\sigma^\prime}$, and $P_d(z)\sim z^{-1}$. In the ER model, $z$ may be interpreted as the age in the birth-to-death process. We confirm the result numerically, as shown in Appendix~\ref{app_age}.

\begin{figure}
	\centering\includegraphics[width=\linewidth]{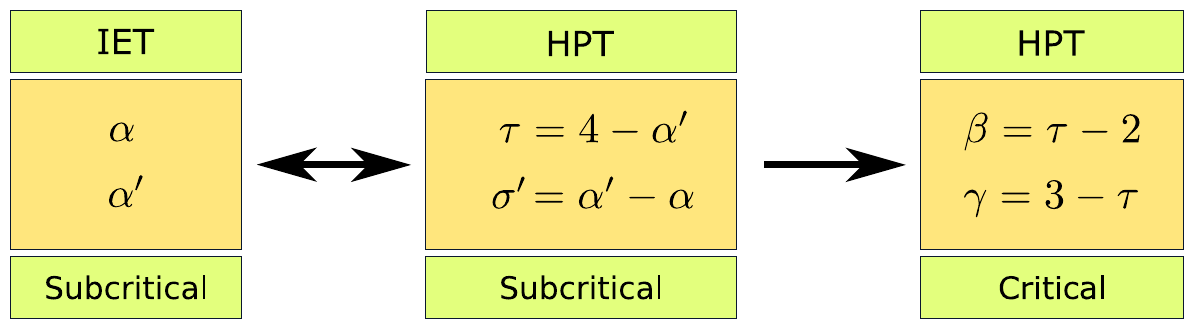} 
	\caption{\textbf{Relationship between the exponents:}
		The IET exponents $\alpha$ and $\alpha^\prime$ determined in subcritical regime are related to the critical exponents $\tau$ and $\sigma^\prime$ of the cluster size distribution. And then, the exponent $\tau$ determines the critical exponents $\beta$ and $\gamma$ of the order parameter and susceptibility.}
	\label{fig:schematic-and-table}
\end{figure}

Next, we calculate the IET distribution $P_{d}(z)$ in $[t_a,t_g]$, which is composed of duration times that terminate in the interval $[t_a,t_g]$. During this short interval, the probability of selecting a cluster of size $s$ may be regarded as $s n_s(t_g)\propto s^{1-\tau} e^{-s/s^*(t_g)}$, where $s^*(t_g)$ is constant, and thus $\sigma^\prime=0$. In this case, $P_d(z)=s \cdot sn_s \left|\partial s/\partial \langle z\rangle \right| \sim z^{-(4-\tau)}$. The exponent $\alpha^{\prime}=4-\tau$ is in reasonable agreement with numerical results for the process $A\to B$, as shown in Appendix~\ref{app:iet}. Note that the two relations, $\alpha=4-(\tau+\sigma^\prime)=1$ for $[0,t_a]$ (and even for $[0,t_g]$) and $\alpha^{\prime}=4-\tau$ for $[t_a,t_g]$, enable us to determine the static exponents $\tau$ and $\sigma^\prime$ for the cluster size distribution, once we have measured the dynamic exponents $\alpha$ and $\alpha^{\prime}$. Furthermore, the exponent $\tau$ determines the critical exponents $\beta$ and $\gamma$ associated with the order parameter and the susceptibility as $\beta=\tau-2$ and $\gamma=3-\tau$, respectively~\cite{cho}. Therefore, the IET exponents $\alpha$ and $\alpha^\prime$ characterize the critical behavior of the HPT induced by cluster coalescence. We summarize these relationships in Fig.~\ref{fig:schematic-and-table}. 

\section{Discussions}
\label{sec:discussion}

We remark that the exponent $\alpha$ is closely related to the exponent of the age distribution of the birth-to-death process in cluster merging kinetics. This age is defined as the interval between two consecutive cluster mergings. The reason is as follows: the exponent $\alpha$ is determined in the early time regime, in which clusters are overall small and thus when a cluster in set $A$ is merged with another cluster regardless of from set $A$ or $B$, it is likely to move to set $B$. Thus, the age and the IET distributions behave similarly, which are checked for ER and $r$-ER models. Next, we check the age distribution of bank mergers and acquisions (M\&A) process based on the data of the US since the year 1900. It shows a power-law decay with a different exponent. The details are presented in Appendix~\ref{app_age}.

Overall, we find that cluster inter-set crossing events occur more frequently as time passes, as shown in Fig.~\ref{fig:event}(b). We examine the number of crossing events, denoted as $N_{\rm event}(t)$, as a function of time by counting the number of nodes that switch from one set to the other at a given time $t$. We argue that $N_{\rm event}(t)\sim  1/\langle z \rangle_s \sim s^*(t) \sim (t_g-t)^{-1}$ with some cutoff. Thus, a burst occurs in the inter-set crossing as the time approaches $t_g$. Numerical data are presented in Appendix~\ref{app_burst}. The burst may signal the upcoming HPT. 

We note that the dynamic critical features of $P_d(z)$ and $N_{\rm event}(t)$ are not limited to the $r$-ER model. Rather, we find similar behaviors in the CMD of the ER model. However, whereas the IET distribution in the $r$-ER model has two distinct power laws in two different time windows, as shown in the insets of Fig.~\ref{fig:event}, the behavior in the second regime, $[t_a, t_g]$, does not occur in the ER model because the transition is continuous. The details are presented in Appendix~\ref{app_waiting}.  

Owing to the suppression effect, medium-sized clusters are accumulated and form a bump in the cluster size distribution. As more time passes, clusters in the bump are merged with each other (or with other small clusters), and then a giant cluster emerges and grows in a Devil's staircase pattern in a short time period, $[t_a,t_c]$~\cite{fraction}. The Devil's staircase pattern shows a power-law jump distribution, $P_m(\Delta m)\sim (\Delta m)^{-\delta}$; the exponent $\delta$ exhibits crossover behavior from $\delta_1=\tau-1$ to $\delta_2=(1+\eta)/\zeta$, where $\eta$ and $\zeta$ are the phenomenological exponents associated with the order parameter. More details are presented in Appendix~\ref{app_noise}. 

\section{Conclusion}
\label{sec:conclusion}
We have investigated the underlying mechanism of the HPT induced by cluster merging dynamics using the $r$-ER model, finding that the IETs between two consecutive set-crossing times have two distributions with power-law decays. This impiies that there exists a self-organized critical (SOC) behavior previously unrecognized in the HPT.  We established a theoretical framework for the HPT, analogous to the conventional percolation theory, and showed that the exponents of the IET distributions determine the critical behavior of the HPT. This SOC behavior is comparative to the critical branching process governing the avalanche dynamics of the HPT in pruning process. 

\section*{Popular summary}
Percolation has long served as a paradigm for diverse phenomena in multidisciplinary area such as network resilience, socio-community formation, and disease contagion. Percolation transition (PT) is known as one of the most robust second-order (continuous) transitions. However, PTs that occur in complex systems often appear in diverse types including first-order (discontinuous) and hybrid transitions. Hybrid percolation transitions (HPTs) exhibit mixed properties of first-order and second-order transitions.

HPTs can occur through pruning and cluster merging processes (CMD). The HPT induced by CMD, denoted as HPT-CMD, is more intrigue and little understood, compared to the HPT in pruning process which is driven by critical branching processes. In this paper, we uncovered the critical behavior underlying the HPT-CMD and established a theory for the HPT-CMD using the so-called restricted percolation model. In this model, clusters are ranked by size and partitioned into small- and large-cluster sets. As the cluster rankings are updated as clusters are merged, clusters may move back and forth across the set boundary. The inter-event times (IETs) between two consecutive crossing times have two distributions with power-law decays, implying a new type of self-organized critical (SOC) behavior. Those are related to the formation and elimination of a bump in the size distribution of finite clusters.

We showed that this SOC behavior characterizes the criticality of the HPT. Moreover, a burst of such crossing events occurs and signals the upcoming transition. We hope this uncovered mechanism can be used for understanding other dynamic critical phenomena with dragon kings.

\begin{acknowledgements}
BK thanks H.~J. Herrmann for useful discussions. This work was supported by the National Research Foundation of Korea (NRF) through Grant Nos. NRF-2014R1A3A2069005 (BK), 2017R1A6A3A11031971 (SY), and 2017R1A6A3A11033971 (DL). 
JP and SY equally contribute to the paper as the first author. 
\end{acknowledgements}

\appendix

\section{Determination of $m_a$}\label{app:m_a}

\begin{figure*}[!h]
	\centering\includegraphics[width=\linewidth]{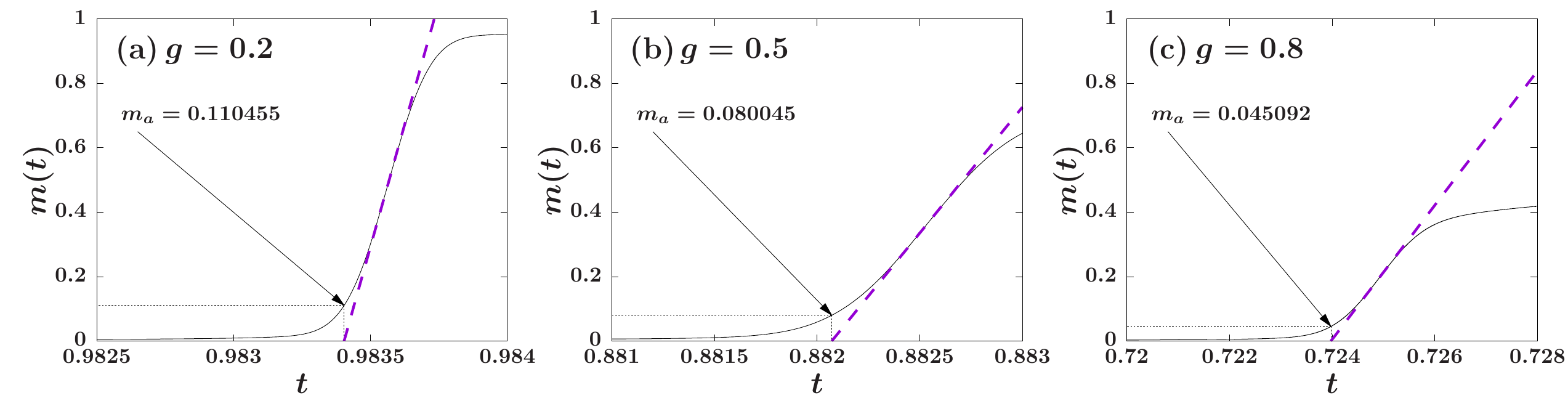}
	\caption{Determination of $m_a$ for different $g$ values. First, we determine $t_a$ as the intercept with the $t$ axis of the tangential line of $m(t)$ at the inflection point of $m(t)$. Then $m_a=m(t_a)$. }
	\label{fig:m_a}
\end{figure*}

First, we determine $t_a$ as the intercept with the $t$ axis of the tangential line of $m(t)$ at the inflection point of $m(t)$ in the region $0 < t < t_c$. Then $m_a=m(t_a)$ (Fig.~\ref{fig:m_a}).

\section{Development of power law region in the cluster size distribution $n_s(t)$ and maximal bump height at $t_a$ }
\label{app:n_s}

\begin{figure*}[!h]
	\centering\includegraphics[width=\linewidth]{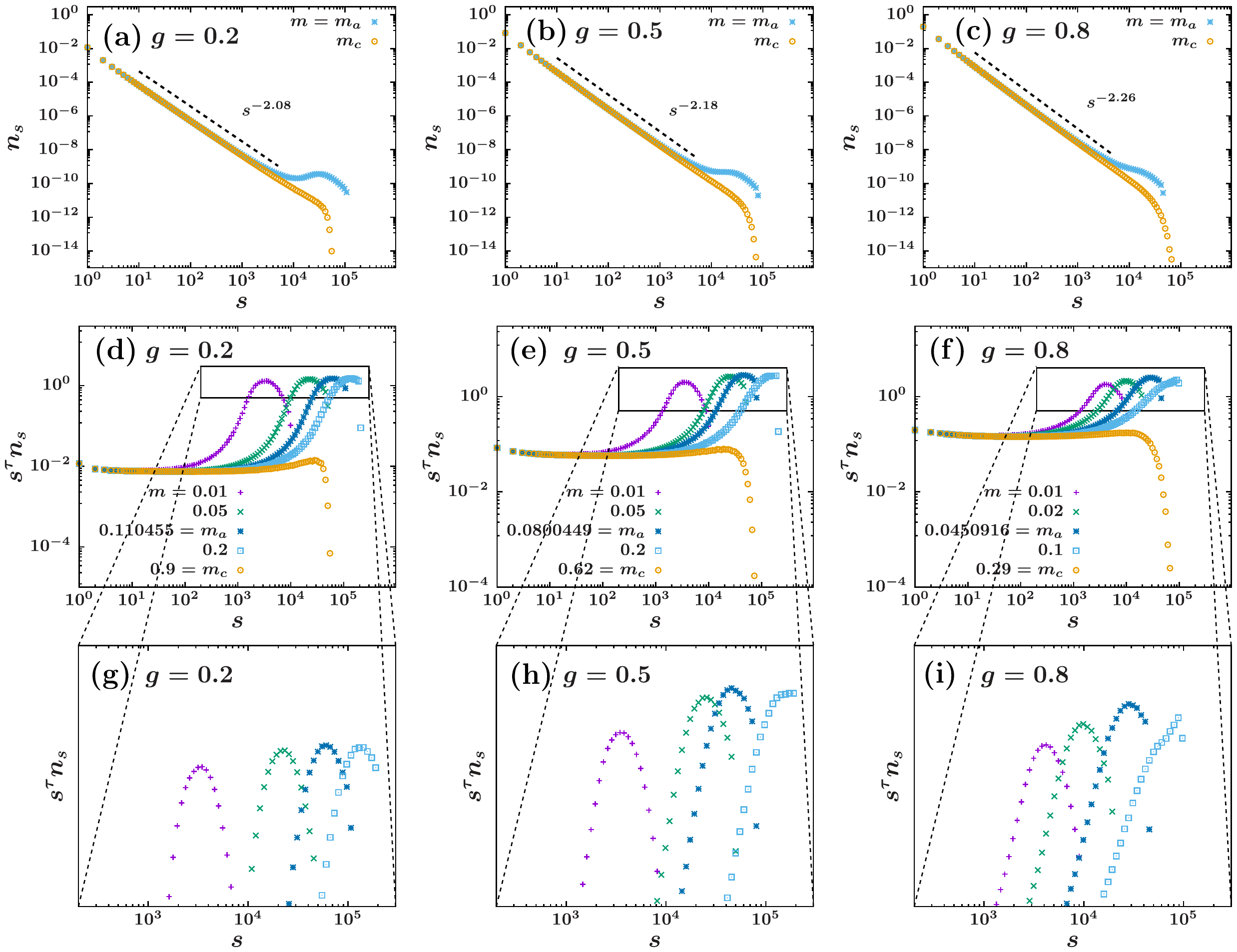}
	\caption{Plot of $n_s$ and $s^{\tau}n_s$ versus $s$ at various values of $m$. (a)$-$(c) Cluster size distribution $n_s(t)$ for $g=0.2,0.5$, and $0.8$ at $t=t_a$ and $t=t_c$, which correspond to $m=m_a$ and $m_c$, respectively. (d)$-$(f) A power-law behavior of $n_s$ is developed in small-$s$ region before $t_a$ and the region is elongated as time proceeds. Bump height seems to be maximum around $m_a$, and as time proceeds further, the right side of the bump is eroded as shown in zoom-in figures (g)$-$(i).}
	\label{fig:tau-and-bump}
\end{figure*}

A power law slope of $n_s$ is slowly developed as the $r$-ER dynamics proceeds. Around the time $t_a$, the height of bump becomes maximum as shown in Fig.~\ref{fig:tau-and-bump}(d)$-$(i), and a giant cluster of size $m_aN\sim O(N)$ begins to grow as in the explosive percolation~\cite{finite}. From then the giant cluster size rapidly increases during the interval $[t_a,t_c]$ of $o(N)$, which signifies a discontinuous transition in the thermodynamic limit. At the same time bump erodes away and a pure power-law cluster size distribution is formed at $t_c$.

\section{The inter-event time distribution} 
\label{app:iet} 

We assign a duration time $z_\ell$ to each node $\ell$. Initially, at $t=0$, $z_\ell=0$ for each node $\ell$. The duration time of each node $\ell$ is updated at each time step as follows: Suppose two nodes $i$ and $j$ are selected from the system under the given rule at a given time $t$. Here unit time is taken as $1/N$. 
\begin{itemize} 
\item[i)] If the two selected nodes belong to the same cluster in set $A$, they are connected unless there is already a link. The duration times of all nodes in the system $\ell \in A,B$ are increased by a unit time $z_\ell \to z_\ell+1/N$. The constituent nodes and cluster sets of the system remain the same. 
\item[ii)] 
If the two nodes belong to two distinct clusters $c_i$ and $c_j$ of sizes $s_i$ and $s_j$, respectively, in set $A$, then they are connected. Accordingly, the two clusters are merged into a new cluster $c_k$ of size $s_k=s_i+s_j$.
\begin{itemize}
\item[ii-1)]If $s_k \le s_A$, then cluster $c_k$ remains in set $A$. The duration times of each node $\ell$ in both sets are updated as $z_\ell \to z_\ell+1/N$. The constituent nodes of both sets remain the same. Note that as a result of this coalescence process, nodes in the same cluster $c_k$ have different duration times.  
\item[ii-2)] If $s_k > s_A$, cluster $c_k$ is moved to $B$, and an approximately same mass of clusters is transferred from $B$ to $A$, in order to retain the mass restrictions of sets $A$ and $B$. Consequently, one or two of the smallest clusters in set $B$ are ejected to $A$. The duration times of all the emigrant nodes are accumulated in the IET distribution $P_d(z)$ and then reset to zero. The duration times of all the other nodes are updated as $z_\ell \to z_\ell+1/N$. $s_A$ may or may not be increased depending on the sizes of the ejected clusters. 
\end{itemize}
\item[iii)] 
If the two selected nodes belong to clusters $c_i \in A$ and $c_j \in B$ of sizes $s_i$ and $s_j$, respectively, the duration times $z_{i}$ of each node $i \in c_i$ are accumulated in the IET distribution $P_{d}(z)$, and then those duration times are reset to zero ($z_i=0$). The duration times of each node $j$ in cluster $c_j$ are updated as $z_j \to z_j+1/N$. The two clusters are merged, generating a cluster $c_k$ of size $s_i+s_j$ in $B$. One of the smallest cluster in $B$, denoted as $c_\ell$, is ejected to $A$, and the duration times of each node in $c_\ell$ are accumulated in $P_{d}(z)$. Then they are reset to zero.  
\end{itemize}

The IET distribution can be constructed by collecting these duration times over all nodes during a given time interval, for instance, $[0,t_g]$. We find that the IET distribution exhibits power-law decay as $P_d(z)\sim z^{-\alpha}$. The exponent $\alpha$ is close to unity for $g=0.2$ and $0.5$; for such small values of $g$, the exponent appears to be insensitive to changes in $g$. However, the exponent $\alpha$ is sensitive to the time interval in which the IET distribution is accumulated. When we accumulate the IET distribution during the interval $[0, t_a]$, $\alpha\approx 1$, but for the interval $[t_a, t_g]$, $\alpha^{\prime} \approx 4-\tau$. The reason is the different dynamics during each interval. Because the IET distribution decays rapidly in the interval $[t_a, t_g]$, the exponent $\alpha$ for the distribution accumulated in the interval $[0,t_g]$ is close to one.   

We consider in detail the behavior of the IET distribution for the duration in set $A$, which is denoted as $P_{d}^{A\to B}(z)$. Let us consider the case in which a cluster $c_0$ of size $s_0$ moves from $B$ to $A$ at time $t_0 > 0$. Cluster $c_0$ grows in $A$ as it is subsequently merged with other clusters, and then it returns to $B$ at time $t_1=t_0+z$, because its size becomes larger than $s_A(t_1)$. $s_A(t)$ is the maximum size of the clusters in set $A$ at time $t$, which divides the system into the two sets $A$ and $B$. It increases with time. Then $s_0$ nodes that belonged to the original cluster $c_0$ at time $t_0$ have duration time $z$ in set $A$. The evolution of cluster $c_0$ is depicted schematically in Fig.~3 of the main text. Each constituent node of a cluster may have a different IET. After sufficient growth in set $A$, a cluster is transferred to set $B$, where its growth is limited. We measure the duration time a node spends in one set before it moves to the other set. The probability $P_{d}^{A\to B}(z)$ that such events happen is calculated as Eq.~\eqref{eq:pd}, where $q_1(s_0;t_0)$ is the probability that a cluster of size $s_0$ in $B$ makes a room for some other immigrant cluster from $A$ and is ejected to $A$ at time $t_0$. The probability $q_1$ is given as  
\begin{align}
q_1(s_0;t_0)&= 1 - \sum_{i+j \le s_0} \frac{in_i(t_0) jn_j(t_0)}{g}. \label{eq:s4} 
\end{align}
Further, $q(s_{t_0+t};t_0+t)$ is the probability that cluster $c_0$ remains in set $A$ with size $s_{t_0+t}$ at time $t_0+t$. For it to remain in set $A$, it is necessary that $u < t$, $s_{t_0+u} \le s_A(t_0+u)$. $q(s_t;t_0+t)$ is calculated as follows:

(i) Let the size of cluster $c_0$ at $t_0+t-1$ be denoted as $s_{t_0+t-1}$. A node in this cluster is selected at $t_0+t-1$ from among all nodes in set $A$ with probability $s_{t_0+t-1}/gN$. Next, a node is selected from a cluster of size $j$ in the entire system, which occurs with probability $jn_j(t_0+t-1)$. At time $t_0+t$, the two nodes are linked, generating a cluster of size $s_{t_0+t}=s_{t_0+t-1}+j$. This size needs to be less than or equal to $s_A(t_0+t)$. This occurs with probability 
\begin{align}\label{c1}
\tilde{q}_1(s_{t_0+t},t_0+t)=2\frac{s_{t_0+t-1}}{gN}\sum_{j=1}^{s_A(t_0+t)-s_{t_0+t-1}}jn_j(t_0+t-1),
\end{align}
where the factor two comes from an alternative way of choosing these clusters.  

(ii) We consider the case that cluster $c_0$ of size $s_{t_0+t-1}$ is not selected at time $t_0+t$, which occurs with probability  
	\begin{align}\label{c2}
	\tilde{q}_2(s_{t_0+t},t_0+t)&=[1-(s_{t_0+t-1}/gN)][1-(s_{t_0+t-1}/N)] \nonumber \\
	&\approx 1-(1+g)\frac{s_{t_0+t-1}}{gN},
	\end{align}
	where we ignore the correction terms of higher order, $O(N^{-2})$. Together, 
	\begin{align}
	q(s_{t_0+t},t_0+t)=\tilde{q}_1(s_{t_0+t},t_0+t)+\tilde{q}_2(s_{t_0+t},t_0+t).
	\end{align}

\begin{figure*}[!h]
	\centering\includegraphics[width=\linewidth]{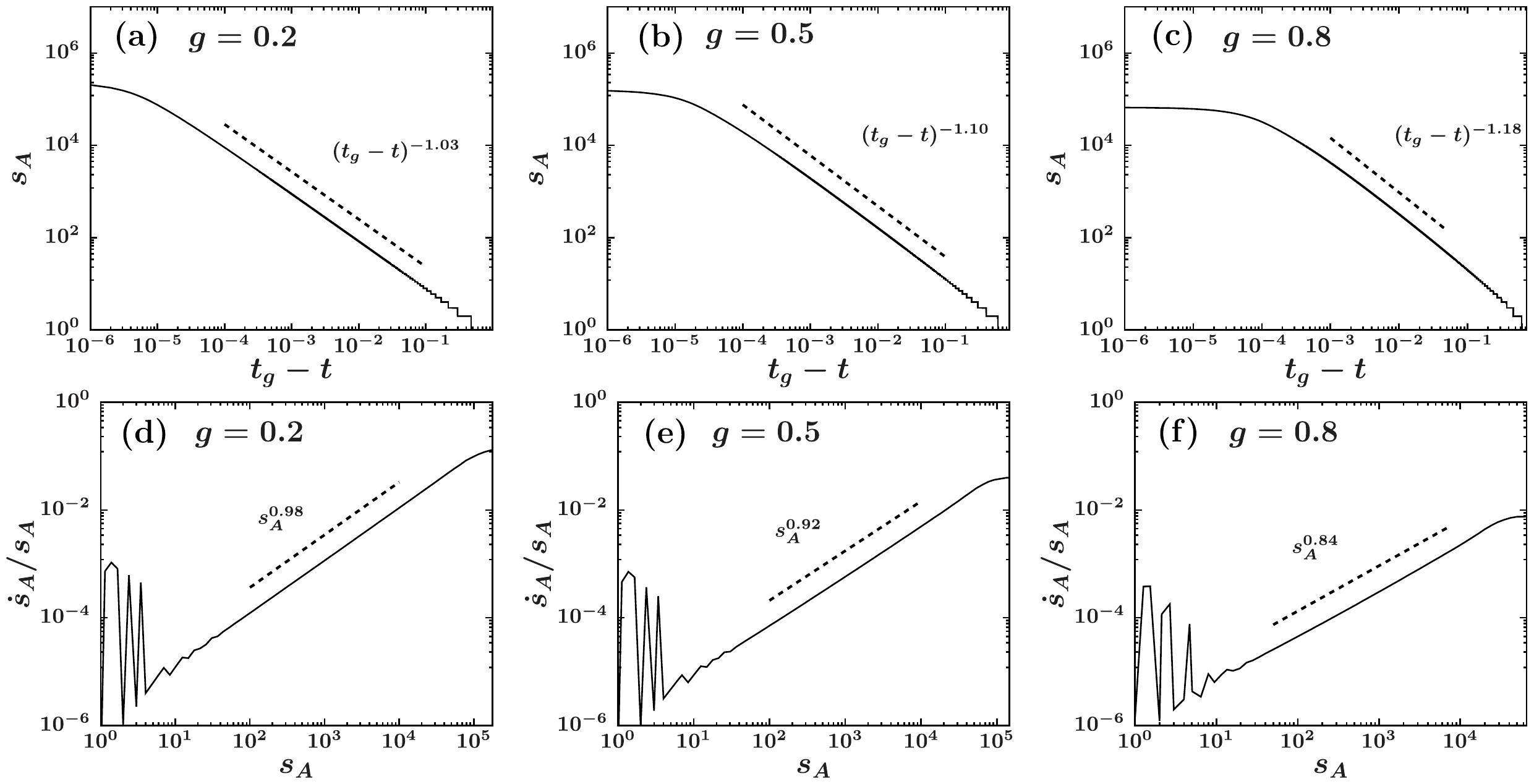}
	\caption{(a)$-$(c) Dynamic scaling of $s_A \sim (t_g-t)^{-1/\sigma^\prime}$ for different $g$ values. (d)$-$(f) $\dot{s}_A / s_A$ versus $s_A$. The curves were averaged over $10^5$ configurations.}
	\label{fig:sigma}
\end{figure*}

The product $\prod_{t=1}^{z-1}q(s_{t_0+t};t_0+t)$ depends on the full history of the size evolution of a cluster that was born with size $s_0$ at time $t_0$ and grew during the duration time $z$.

To calculate Eq.~\eqref{eq:pd}, we consider the largest cluster size $s_A(t)$ at time $t$ averaged over different realizations. We find numerically, as shown in Fig.~\ref{fig:sigma}(a)$-$(c), that 
\begin{align}
s_A(t)\sim (t_g-t)^{-1/\sigma^\prime}h\Big(\frac{t_g-t_a}{t_g-t}\Big),  
\label{eq:sigma}
\end{align}
where $h(x)$ is a scaling function, defined as $h(x)\sim s^{-1/\sigma^\prime}$ for $x < 1$ and may be regarded as constant for $x > 1$ and $\sigma^\prime$ depends on the control parameter $g$.
This gives an alternative form in the region $t\in [0,t_a]$,
\begin{align}
\frac{s_A}{\dot{s}_A}={\sigma^\prime(t_g-t)} \sim \sigma^\prime {s_A^{-\sigma^\prime}}.
\end{align}

\begin{figure*}[h]
	\centering\includegraphics[width=\linewidth]{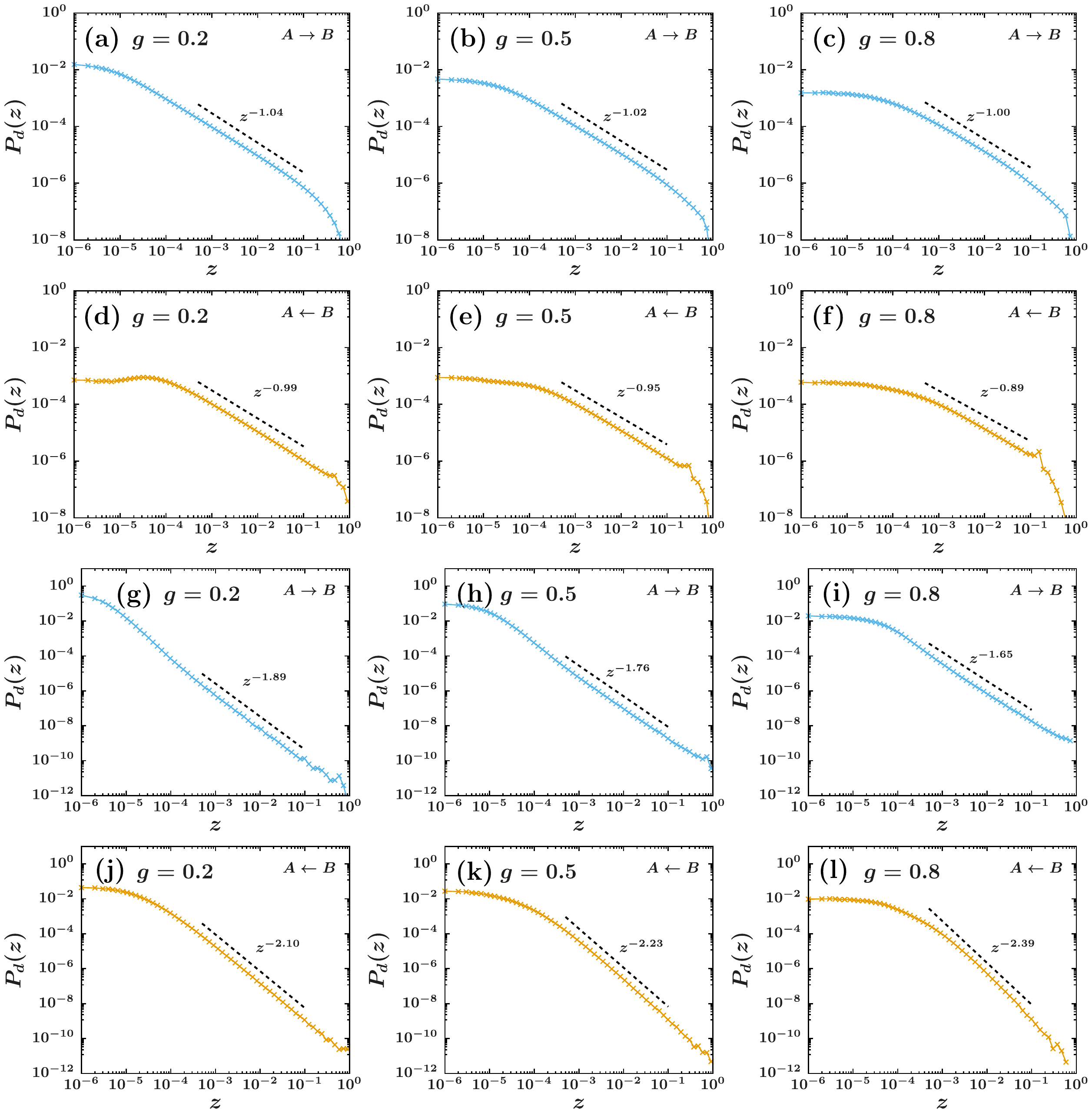} 
	\caption{ IET distributions accumulated in the time windows (a)$-$(f) $[0,t_a]$ and (g)$-$(L) $[t_a,t_g]$ versus duration time $z$. The exponent $\alpha$ is modified only slightly, depending on the direction of switching: (a)$-$(c),~(g)$-$(i) $P_d^{A\rightarrow B}(z)$ and (d)$-$(f),~(J)$-$(L) $P_d^{A\leftarrow B}(z)$}
	\label{fig:interevent}
\end{figure*}

To calculate $q_1$ and $q$, we need the explicit form of the cluster size distribution $n_s(t)$. Even though the cluster size distribution includes a bump in the medium-size region centered at $s_A(t)$, we ignore the contribution of the bump to the probabilities $q$ and $q_1$ without losing the essence. Moreover, we take the power-law region up to $s_A(t)$. That is, $n_s(t)\sim s^{-\tau}$ for $s \le s_A(t)$, where $\tau$ depends on $g$~(Fig.~\ref{fig:tau-and-bump}(a)$-$(c)). Then, from the normalization $\sum_{s=1}^{s_A} sn(s,t)=g$, it follows that
\begin{align}
n_s(t) = \frac{g(\tau-2)}{1-s_A^{2-\tau}(t)}s^{-\tau}\equiv C_As^{-\tau},
\end{align}
where $C_A$ is the normalization factor of the cluster size distribution in set $A$. 
Now we introduce the probability of having selected two nodes in set $A$:
\begin{align}
w_0(s,t)&\equiv \frac{1}{g}\sum_{i+j\le s} in_i(t)jn_j(t)\nonumber \\ 
&\simeq \frac{1}{g} \int_1^{s}di~in_i(t) M_1(s-i,t),
\end{align}
where $M_1(s,t)=\sum_{j=1}^{s} j n_j(t)$ is the accumulated mass function in set $A$, which satisfies the normalization $M_1(s_A,t)=g$. In particular, 
when $s\to s_0\approx s_A$ in $q_1(s_0,t_0)$, 
\begin{align}
w_0(s,t)&\approx \frac{1}{g} \int_1^{s}di~in_i(t) M_1(s_A-i,t), \\
w_1(s,t)&\equiv\frac{2s}{gN} \sum_{j=1}^{s_A(t)-s} jn_j(t) \xrightarrow{s\to s_0} \frac{2}{Nn_s}\frac{dw_0(s)}{ds}.
\end{align}
Then,  
\begin{align}
q_1(s_0,t_0)=1- w_0(s_0,t_0),
\end{align}
and 
\begin{align}
q(s,t)&={\tilde q}_1(s,t) + {\tilde q}_2(s,t)\nonumber \\
&=\frac{2s^{\tau}}{N C_A}\frac{dw_0(s)}{ds} + \bigg[1-\frac{(1+g)s}{gN}\bigg].
\end{align}
Next, $w_0(s)$ is calculated as 
\begin{align}
w_0(s)&\approx \frac{g}{\big(1-s_A^{2-\tau}\big)^2}
\bigg[1 - 2s^{2-\tau} + C_{\tau} s^{4-2\tau} \nonumber \\ &-\frac{(\tau-2)(4-\tau)}{3-\tau}s^{1-\tau}+(\tau-2)(3-\tau)s^{-\tau}+\cdots\bigg],
\end{align}
where 
\begin{align}
C_\tau\equiv\frac{\Gamma(3-\tau)\Gamma(3-\tau)}{\Gamma(5-2\tau)}= \frac{\sqrt{\pi}4^{\tau-2}\Gamma(3-\tau)}{\Gamma(\frac{5}{2}-\tau)}.
\end{align}
Then $q_1$ and $q$ are approximated to be stationary in time and only kept up to the leading orders:
\begin{align}
q_1(s) &\approx 1-g , \label{eq:q1} \\
q(s)&\approx 1-\frac{1+5g}{gN}s + \frac{4C_{\tau}}{N}s^{3-\tau} \nonumber \\ 
&- \frac{2}{N}\frac{(\tau-1)(4-\tau)}{3-\tau} + O\big(s^{-1}, s_A^{2-\tau}\big). \label{eq:q}
\end{align}
Then the IET distribution in Eq.~\eqref{eq:pd} is obtained as
\begin{widetext}
\begin{align}
P_{d}^{A\to B}(z) &\simeq \int_{1}^{s_A(t_g)} ds_A(t_0) ~ \frac{1}{ds_A/dt_0}s_0 q_1(s_0;t_0)\delta\big(s_0-s_A(t_0)\big)\bigg[\prod_{t=1/N}^{z-1/N} q(s_{t+t_0};t_0+t)\bigg]\big[1-q(s_{t_0+z};t_0+z)\big] \label{eq:s20} \\
&\simeq \sigma^\prime\int_{1}^{s_A(t_g)} ds_0 ~ s_0^{-\sigma^\prime} q_1(s_0;t_0)\bigg[\prod_{t=1/N}^{z-1/N} q(s_{t_0+t};t_0+t)\bigg]\big[1-q(s_{t_0+z};t_0+z)\big], \quad\quad  {\rm for} \quad z > 1/N  \label{eq:s21} \\  
&\sim \frac{(1-g)\kappa\Gamma(2-\sigma^\prime)N^{\sigma^\prime-2}}{\alpha \big(\log\frac{1}{1-A}\big)^{2-\sigma^\prime}}\bigg[z^{\sigma^\prime-2}+\frac{B}{1-A}~\frac{\Gamma(5-\sigma^\prime-\tau)N^{\tau-2}}{\Gamma(2-\sigma^\prime)}\frac{1}{\big(\log\frac{1}{1-A}\big)^{3-\tau}}~z^{\sigma^\prime+\tau-4}\bigg]e^{-z/z_c} \label{eq:s22} \\
&\sim z^{-(4-\tau-\sigma^\prime)}e^{-z/z_c},
\label{eq:s23}
\end{align} 
\end{widetext}
where $\kappa\equiv \dot{s}_t/s_t$, $A\equiv{(1+5g)}/({gN})$, $B\equiv {4C_\tau}/{N}$, and $z_c=(3-\tau)/[2(\tau-1)(4-\tau)]$. We remark that in the second line to third line, we used the approximations~\eqref{eq:q1} and \eqref{eq:q}. Then it is noticed that the \eqref{eq:s21} corresponds to an integral representation of incomplete beta function. This integral can asymptotically evaluated as in~\eqref{eq:s22} for sufficiently large upperbound of the integral $s_A(t_g)\gg 1$. Moreover, in the limit of large interval $z$ we have the asymptotic expression~\eqref{eq:s23}. Therefore, we finally obtain the relation
\begin{align}
P_d^{A\to B}(z)\sim z^{-\alpha} \sim z^{-(4-\tau-\sigma^\prime)}. \label{eq:s24}
\end{align}
In the main text, we argue that $\tau+\sigma^\prime=3$. Thus, $\alpha=1$. In Fig.~\ref{fig:interevent}, we show that this analytic result is consistent with numerical data. 

\section{Waiting time distribution}\label{app_waiting}

Here we calculate the IET distribution $P_{d}(z)$ by an alternative method. We consider the mean waiting time steps $Z\equiv zN$ of a cluster of size $s$ in set $A$ until it is chosen. 
\begin{align}
\langle Z\rangle &= \int_0^\infty dZ~ Z\Big(1-\frac{s}{N}\Big)^Z\Big(1-\frac{s}{gN}\Big)^{Z} \nonumber \\
&\cdot \Big[\frac{s}{gN}\Big(1-\frac{s}{N}\Big) + \frac{s}{N}\Big(1-\frac{s}{gN}\Big) \Big] \\
&\simeq \frac{(1+g)s}{gN}\int_0^\infty dZ~ Z e^{-\frac{(1+g)s}{gN}Z} \\
&=\frac{gN}{(1+g)s}.
\end{align}

Using this property, we calculate the IET distribution $P_d^{A\to B}(z)$ as 
\begin{align}
P_d(z)\sim s\cdot sn_s\left|\frac{\partial s}{\partial z}\right|.
\end{align}
During the entire period $[0,t_g]$, the mean cluster density $\langle n_s(t)\rangle =(1/t_g)\int_0^{t_g}dt s^{-\tau}f\left(s/s^*(t)\right)$, where $s^*(t)\sim (t_c-t)^{-1/\sigma^\prime}$.
$s^*(t)$ may be replaced by $s_A(t) \sim (t_g-t)^{-1/\sigma^\prime}$. Dimensional analysis obtains
\begin{align}
\langle n_s(t) &\rangle = \frac{1}{t_g}\int_0^{t_g} dt s^{-\tau}f\big(s(t_g-t)^{1/\sigma^\prime}\big)\\
&= \frac{s^{-\tau}}{t_g}\int_0^{t_g} dt f\big(st^{1/\sigma^\prime}\big)\\
&= \frac{s^{-\tau-\sigma^\prime}}{t_g}\int_0^{st_g^{1/\sigma}} dy~ y^{\sigma^\prime-1}f(y)\\
& \sim \frac{s^{-\tau-\sigma^\prime}}{t_g}\int_0^{\infty} dy~ y^{\sigma^\prime-1}f(y),
\end{align}
where we asymptotically calculate the integral in the last line, using the assumption that the function $f(x)$ decays exponentially at large $y$. Therefore, 
\begin{align}
P_d(z) \sim z^{-(4-\tau-\sigma^\prime)}.
\end{align}

During the short time window $[t_a,t_g]$, a cluster of size $s$ is selected with a probability proportional to $s n_s(t_g)\propto s^{1-\tau} e^{-s/s^*(t_g)}$ at each time step. Nodewise counting in the IET distribution measurement additionally introduces a multiplication factor of $s$. Therefore, 
\begin{align}
P_d(z)= s \cdot sn_s \left|\frac{\partial s}{\partial z}\right| \sim s^{4-\tau} \sim z^{-(4-\tau)}\equiv z^{-\alpha^{\prime}}.
\end{align}
This result is in good agreement with the numerical results shown in Fig.~\ref{fig:interevent}(g)$-$(i). 

\section{The age distributions in the birth-death process of ER and $r$-ER networks}
\label{app_age}

\begin{figure*}[!h]
	\centering\includegraphics[width=\linewidth]{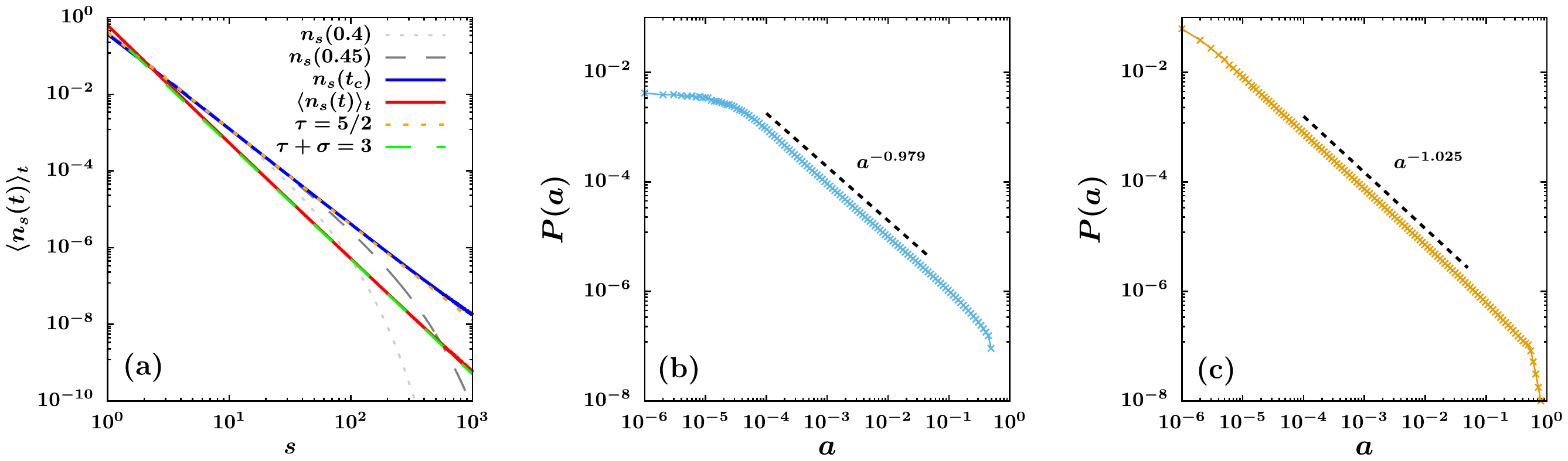}
	\caption{In the $g\rightarrow 1$ limit, the $r$-ER model reduces to the ER model which has $n_s(t_c)\sim s^{-\tau}$ with $\tau=5/2$ at $t_c=0.5$ and $\sigma=\sigma^\prime=1/2$. By solving Smoluchowski equation of the ER model we find that (a) $\langle n_s(t) \rangle_t = \frac{1}{t_c}\int_{0}^{t_c}n_s(t)dt \sim \frac{1}{t_c}\int_{0}^{t_c}dt s^{-\tau} f(s(t_c-t)^{1/\sigma}) \sim s^{-\tau-\sigma}\sim s^{-3}$ is satisfied. Furthermore, regarding cluster merging of the ER model as a birth-death process we find that (b) the age distribution $P(a)$ exhibits power-law decay with exponent close to $1$. In the limit $g\rightarrow 1$, waiting time $z$ can be interpreted as age $a$. In (b) we accumulate the age distribution of nodes in the merging clusters during $[0,t_c]$ and averaged over $10^4$ independent configurations of finite size $N=10^5$ of ER networks. (c) Distribution of merging intervals $P(a)$ in the $r$-ER model. $g=0.2$ and $N=10^4$ are taken. Data are collected during the time window $[0,t_g]$ and from $10^3$ samples.}
	\label{fig:age}
\end{figure*}

Finally, we remark that in the limit $g\to 1$, the $r$-ER model reduces to an ordinary ER model with $\tau=5/2, \sigma=\sigma^\prime=1/2$. So it is easily checked that the universal relation $\tau+\sigma=3$ is also satisfied in this limited case. By solving the rate equation numerically, we find that $\langle n_s(t)\rangle_t\equiv\frac{1}{t_c}\int_0^{t_c}n_s(t)dt\sim \frac{1}{t_c}\int_{0}^{t_c}dt s^{-\tau} f(s(t_c-t)^{1/\sigma}) \sim s^{-\tau-\sigma}$ as shown in the Fig.~\ref{fig:age}(a). In the ER cluster aggregation process, waiting time roughly corresponds to the birth-to-death age. We find instead of \eqref{eq:s24} that the age distribution of nodes follows a heavy tailed distribution $P(a)\sim a^{-(4-\tau-\sigma)} \sim a^{-1}$. The age distribution was accumulated during the whole subcritical regime $t\in[0,0.5]$, selectively for those nodes in the merging clusters (Fig.~\ref{fig:age}(b)). The waiting time $z$ of $r$-ER model is absent in the ER model, but the waiting time can be conceptually extrapolated to the age $a$ in the limit $g\to 1$.
Actually, the age distribution can be also measured for the $r$-ER model. It reveals that the distribution has the same exponent value, as shown in the Fig.~\ref{fig:age}(c). We remark that the exponent $\alpha$ is determined in the early time regime, in which clusters are overall small and thus when a cluster in set $A$ is merged with another cluster regardless of from set $A$ or $B$, it is likely to move to set $B$. Thus, the age distribution in ER and $r$-ER models and the IET distributions in $r$-ER model behave similarly.

\begin{figure*}[!h]
\centering\includegraphics[width=0.7\linewidth]{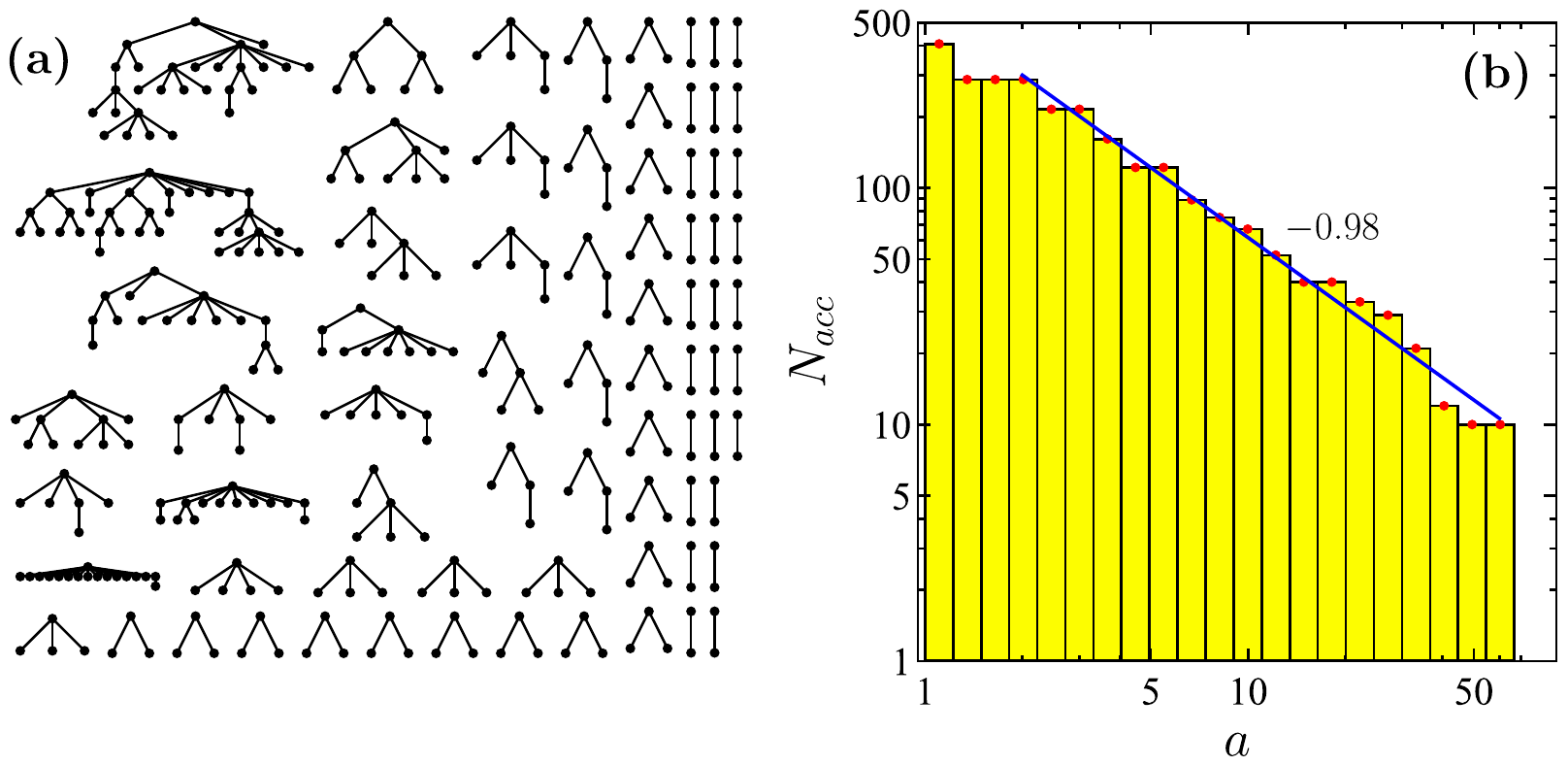}
\caption{(a) Chart of major bank company mergers in the United states since 1900~\cite{bank}. (b) Accumulated histogram of the weighted number of bank mergers $N_{acc}$ versus age $a$ in the unit of a year. Age is the length of an inter-merging period. Ages of the merger and acquirer are collected after each mergers and acqusitions (M\&A) event, and their occurences are weighted by a representative size, defined as the number of leaf banks under each entity.}
\label{fig:bank}
\end{figure*}

Mergers and acquisitions (M\&A) of corporations is a birth-death process which we view as a cluster merging process. Even though M\&A of corporations are caused by many different reasons such as economic shocks and manager's overconfidence, here we apply the random selection rule as used in ER and $r$-ER models as a minimal model. Moreover, we argue that $r$-ER model is more appropriate than ER model, because two big banks are hardly merged. For simplicity, we regard the size of a merged entity as the number of original and acquired corporations. This point may be also inappropriate to reality. Nevertheless, we obtain a heavy-tailed distribution: Fig.~\ref{fig:bank}(a) shows the chart of major banking company mergers in the United States since the year 1900~\cite{bank}. Each link represents a merging event and the roots of the trees represent current major banks in the United States. In Fig.~\ref{fig:bank}(b), we analyzed the distribution of the interval between two consecutive mergings, i.e. age $a$, which indicates how long the bank has been sustained without M\&A. As in the birth-to-death age distributions of $r$-ER and ER networks, the distibution is accumulated on a node base. Each leaf node is regarded as a unit size. The size of an intermediary node is defined as the number of leaf nodes that it has acquired. We present an accumulated histogram in the direction of right to left. The distribution is heavy tailed and has an exponent near unity. This result implies that the inter-merging age distribution of the major bank company mergers has a power-law exponent near two.

\section{Burst}\label{app_burst}

\begin{figure*}[!h]
\centering\includegraphics[width=0.8\linewidth]{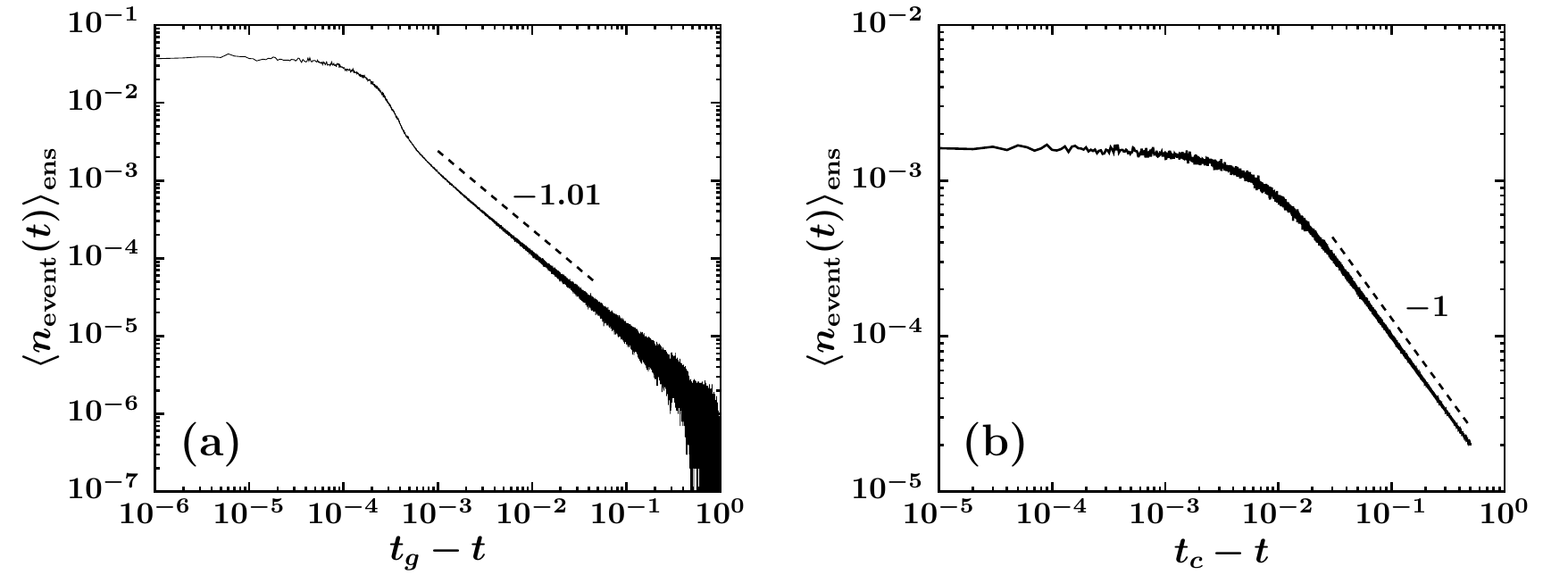}
\caption{Plots of $n_{\rm events}(t)\equiv N_{\rm events}(t)/N$ vs. $t_g-t$ for (a) the $r$-ER model with $g=0.2$ and (b) the ER model. Each figure shows an ensemble averaged curve over $10^3$ samples. Both cases show a power-law behavior with exponent one.}
\label{fig:nevent}
\end{figure*}

The number of nodes that inter-cross from one set to the other increases as time proceeds. We denote it as $N_{\rm event}(t)$. Fig.~\ref{fig:nevent}(a) shows an intermediate range of time where the ensemble averaged number of events follows a power law in time, $\langle N_{\rm event} \rangle_{ens}\sim (t_c-t)^{-1}$. Fig.~\ref{fig:nevent}(b) shows a corresponding power law observed in the ER model in a range of subcritical time, where in the ER model $N_{\rm event}(t)$ is defined as total number of nodes in the two clusters that merge at time $t$. Such a power law behavior can be understood as follows.
\begin{align}
N_{\rm event}(t) &\sim 2\langle s(t)\rangle \sim 2\int ds s\cdot sn_s(t) \label{eq:nevent}\\
&\sim 2\int ds s^{2-\tau} e^{-s(t_c-t)^{1/\sigma^\prime}}\\
&\sim 2(t_c-t)^{(\tau-3)/\sigma^\prime} \int dz z^{2-\tau}e^{-z} \\
&\sim \frac{2\,\Gamma(3-\tau)}{(t_c-t)^{-(3-\tau)/\sigma^\prime}}.
\end{align}
For ER model $\tau=5/2$ and $\sigma^\prime=1/2$. Thus, the exponent is unity. For $r$-ER model with $g=0.2$ $\tau=2.13$ and $1/\sigma=1.033$. Thus we also find an exponent close to unity. We remark that the right hand side of Eq.~\eqref{eq:nevent} corresponds to the definition of the susceptibility in the subcritical regime,
\begin{align}
\chi^-(t) \equiv \sum_s s^2n_s(t) \sim(t_c-t)^{-\gamma^\prime},
\end{align}
where $\gamma^\prime$ has been found numerically~\cite{cho} to be close to unity. Here we have derived the formula for $\gamma^\prime$ with $\gamma^\prime=(3-\tau)/{\sigma}$,  which is indeed close to unity.

\section{Crackling noise}\label{app_noise}
  
\begin{figure*}[!h]
\centering\includegraphics[width=\linewidth]{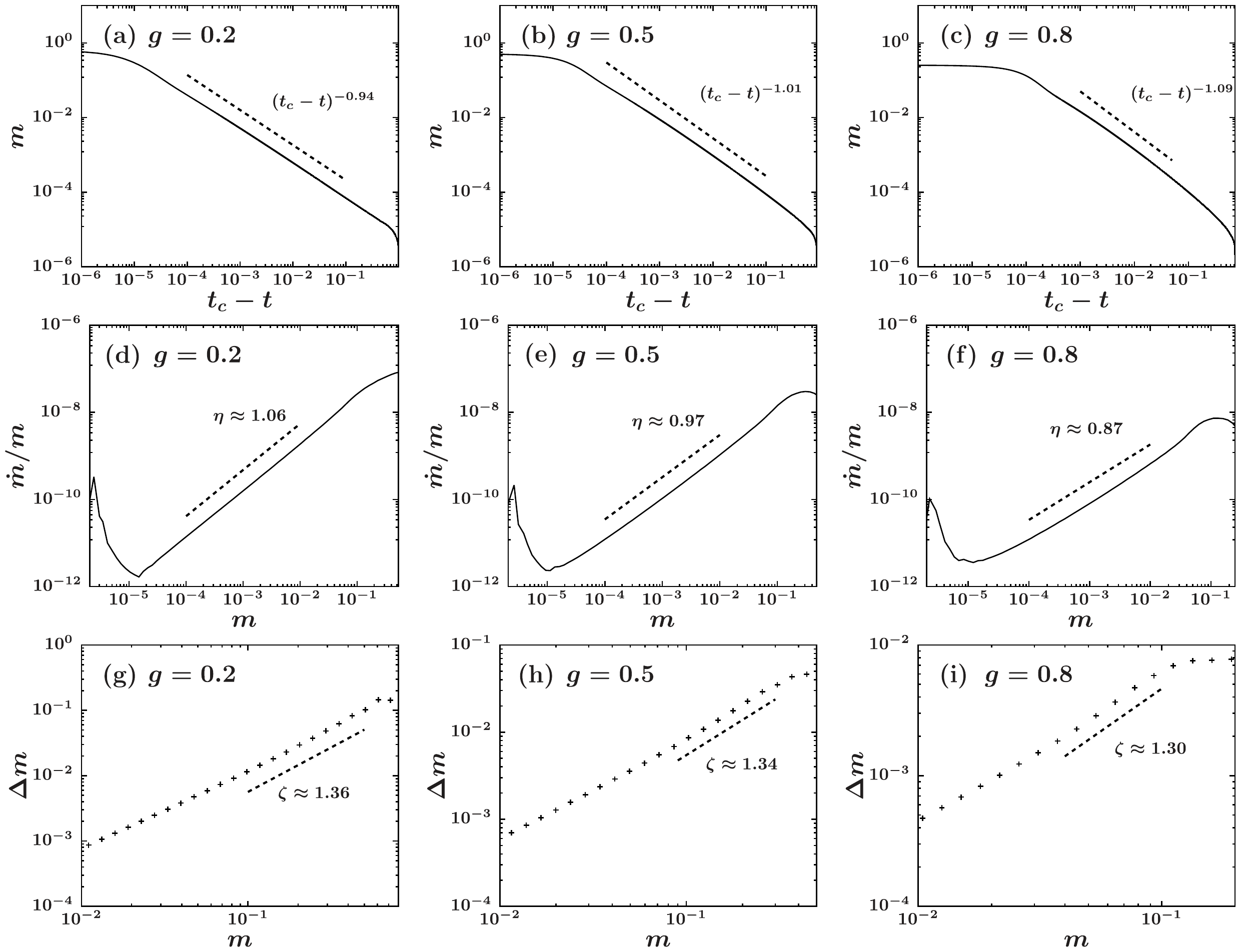}
\caption{(a)$-$(c) Plot of $m(t)$ versus $t_c-t$ to check $m(t)\sim (t_c-t)^{-1/\eta}$. (d)$-$(f) Plot of $\dot{m}/m$ versus $m$. (g)$-$(i) Plot of $\Delta m$ versus $m$. The curves were averaged over $10^5$ configurations. $m(t)$ appears to follow a power law as $m(t)\sim(t_c-t)^{-1/\eta}$ in the intermediate range of $m$, which corresponds to the jump region. The slope in (d)$-$(f) corresponds to the exponent $\eta$. In (g)$-$(i), the data points averaged over $10^5$ configurations are log-binned. The exponent $\zeta$ is measured in the intermediate range of $m$, in which the order parameter increases dramatically. (a) $\zeta =1.36$, (b) $\zeta=1.34$, and (c) $\zeta=1.30$ for $g=0.2,0.5,0.8$, respectively.}
\label{fig:eta-and-zeta}
\end{figure*}

In this section, we consider the evolution of the largest cluster in the system. At each time $t$, we measure $m(t)$. For a single configuration, it increases discontinuously like a staircase with irregular steps. We are interested in the distributions of the jump sizes and widths of the staircase. To calculate these distributions, we consider the continuous order parameter $m(t)$ after taking the ensemble average and find the following empirical scaling relation in the intermediate range of $m$.
\begin{align}
m(t)\sim (t_c-t)^{-1/\eta}, \label{eq:eta}
\end{align}
and thus
\begin{align}
\frac{m(t)}{{\dot m(t)}}={\eta(t_c-t)}\sim \eta m^{-\eta}.
\end{align}
This power-law behavior is shown in Fig.~\ref{fig:eta-and-zeta}(d)$-$(f). \\

\begin{figure*}[!h]
\centering\includegraphics[width=\linewidth]{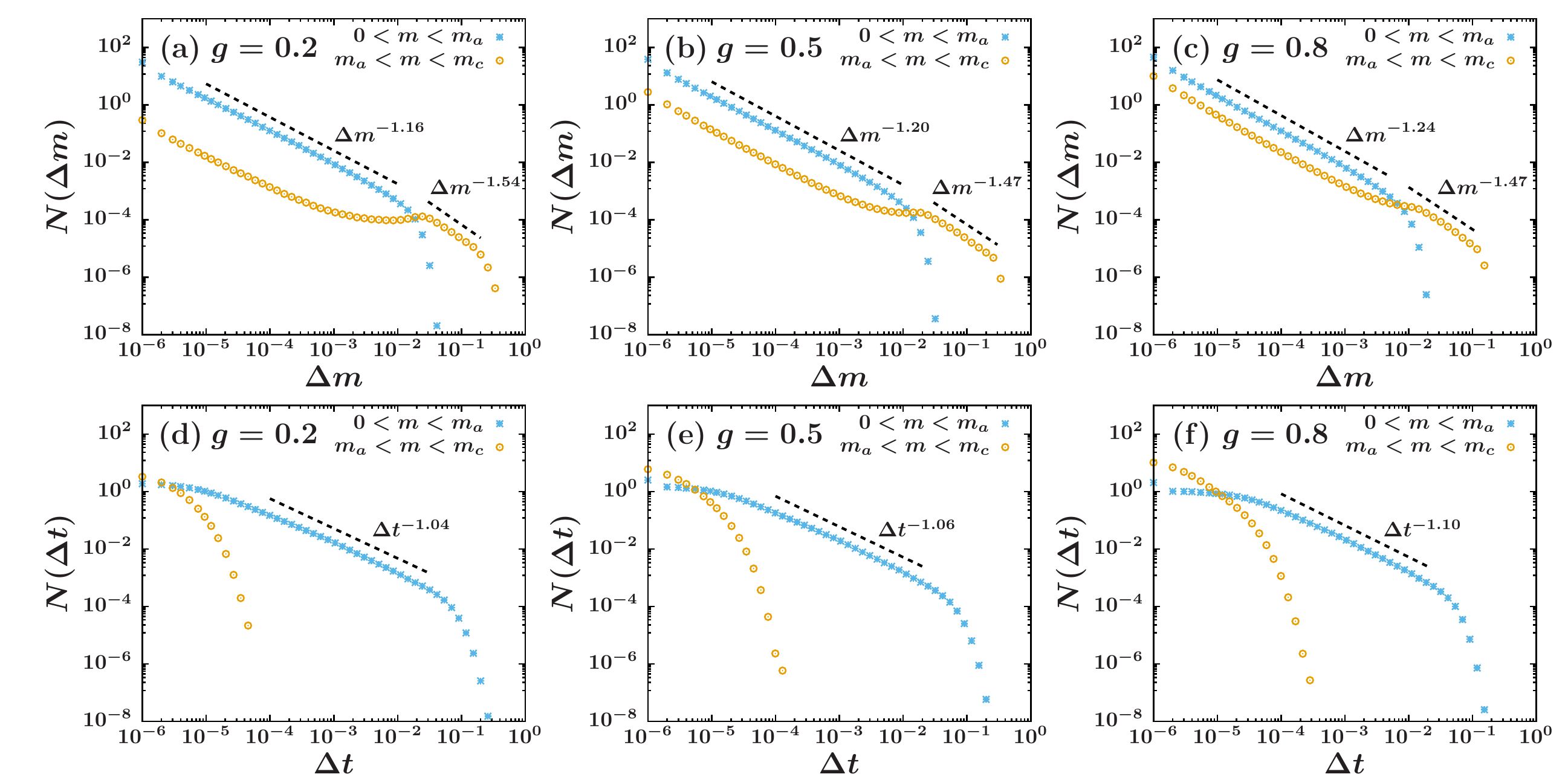}
\caption{Distributions of (a)$-$(c) step height $\Delta m$ and (d)$-$(f) width $\Delta t$. The exponent of $N(\Delta t)$ seems to be independent of $g$.}
\label{fig:jump}
\end{figure*}

Next, we consider the width distribution of the staircase. This can be calculated as follows:
\begin{align}
&P_w(\Delta t;t_c)\nonumber \\ 
&= \int_{0}^{t_c} dt_0 q(m_{t_0})\prod_{t=t_0+1/N}^{t_0+\Delta t-1/N} q(m_t) \big[1-q(m_{t_0+\Delta t})\big]\\
&= \int_{1/N}^{m_{t_c}} \frac{dm}{\dot{m}}\Big|_{t=t_0}m_{t_0}\prod_{t-t_0=1/N}^{\Delta t-1} (1-m_{t_0})m_{t_0}\\
&\approx \eta \int_{1/N}^{m_{t_c}} dm~ m^{1-\eta}e^{-m N(\Delta t-1)}\\
&\simeq \frac{\eta \Gamma(2-\eta)}{(N\Delta t)^{2-\eta}},
\label{eq:deltat}
\end{align}
where $q(m_t)$ denotes the probability that a giant cluster at time $t$ has the same size as it did in the previous time step. We use a normalized time and normalized size, so the index $t$ in the product should be increased by $1/N$, and the size of the giant cluster at $t_c$ is $N m_{t_c}\sim O(N)\rightarrow \infty$ in the thermodynamic limit. Numerically, we obtain $\eta \approx 1$, and thus $P_w(\Delta t)\sim 1/\Delta t$, which is in agreement with the simulation result in Fig.~\ref{fig:jump}(d)$-$(f).  

Next, we consider the portion of accumulation in $P_w(\Delta t)$ that is contributed during some limited time window $[t_a,t_c]$. This marginal distribution decays exponentially, in agreement with the simulation results presented in Fig.~\ref{fig:jump}(d)$-$(f) (orange symbols).
\begin{align}
\frac{dP_w (\Delta t;t_c)}{dt_c} &= \frac{d}{dt_c}\eta \int_{1/N}^{m_{t_c}} dm ~m^{1-\eta}e^{-m N\Delta t}\\
&= \dot{m}_{t_c}\frac{d}{dm_{t_c}} \eta \int_{1/N}^{m_{t_c}} dm ~m^{1-\eta}e^{-m N\Delta t}\\
&= \dot{m}_{t_c} \eta ~ m_{t_c}^{1-\eta}e^{- m_{t_c}N\Delta t}\\
&= \frac{1}{\eta} m_{t_c}^{1+\eta} \eta~m_{t_c}^{1-\eta}e^{-m_{t_c}N\Delta t}\\
&= m_{t_c}^2 e^{-m_{t_c}N\Delta t}.
\end{align}
Moreover, we also consider the jump distribution of the staircase. For intermediate values of $\Delta m$ after $t_a$, the stair heights correspond to the giant cluster size increment. We find that the distribution of the giant cluster size increment $\Delta m$ exhibits crossover behavior. Specifically, it has a power-law tail with an exponent of approximately $1.42$ to $1.54$ (Fig.~\ref{fig:jump}(a)$-$(c)). The bump around $\Delta m \approx 10^{-2}$ corresponds to the abundance of clusters of size $S_A(t_a)$ at time $t_a$.

From Eq. \eqref{eq:eta}, $\dot{m} \sim \eta^{-1} m^{1+\eta}$. From Fig.~\ref{fig:jump}(d)$-$(f) (orange data points), we find that $P_{\Delta t}$ decays exponentially. Thus, we regard $P_{\Delta t}$ as a constant and obtain
\begin{align}
P_{\Delta m} &= \frac{P_{\Delta t}}{\Delta m/\Delta t} \sim \frac{P_{\Delta t}}{\dot{m}} \nonumber \\ 
&\sim m^{-(1+\eta)}\sim (\Delta m)^{-(1+\eta)/\zeta}\nonumber \\ 
&\equiv (\Delta m)^{-\delta},
\end{align}
where $\zeta$ is defined as $\Delta m \sim m^\zeta$ at $[t_c^-,t_c]$. In Fig.~\ref{fig:eta-and-zeta}(g)$-$(i), the value of $\zeta(g)$ is measured to be approximately $1.33$, even though it weakly depends on $g$. Thus, it is expected that 
$$P_{\Delta m}\sim (\Delta m)^{-1.5}.$$
It is worth noting previous results: In an avalanche process such as $k$-core, and the Bak--Tang--Wiesenfeld model on ER networks, the duration of the avalanche $P_\ell \sim \ell^{-1}$, and the size of the avalanche scales as $P_s \sim s^{-3/2}$. From our result, we find that this is also the case. In fractional percolation, multiplicative growth of the jump size has induced $P_s\sim s^{-1}$, and this was interpreted as indicating that the non-self-averaging property may lead to an exponent that is larger than unity. Further, in Barkhausen percolation, the exponent was $1.5$. However, the previously studied underlying mechanism of crackling noise during the avalanche process is not applicable to our model, which is based on dichotomous cluster merging processes.

\section{Table}\label{table}

The numerical values of various exponents $\tau$, $\sigma$, $\alpha$, $\delta$, $\eta$, and $\zeta$ are listed for  several values of $g$. 
\begin{table*}[!h]
\centering
\begin{tabular}{cccccccccc}
\hline
~$g$~&~$\tau^*$~&~$\tau$~~&~$\sigma^\prime$~~&$\alpha$~~&$\alpha^{\prime}$~~&$\delta_1$~~&$\delta_2$~~&$\eta$~~&$\zeta$\\
\hline
~0.2~ &~2.13~ &~2.08~ &~0.974~~ &0.99~~ &2.10~~ &1.16~~ &1.54~~ &1.06~~ &1.36~~\\
~0.5~ &~2.26~ &~2.18~ &~0.934~~ &0.95~~ &2.10~~ &1.20~~ &1.44~~ &0.97~~ &1.34~~\\
~0.8~ &~2.35~ &~2.25~ &~0.906~~ &0.88~~ &1.94~~ &1.24~~ &1.42~~ &0.87~~ &1.30~~\\
\hline
\end{tabular}
\caption{$g$ is the control parameter representing the fraction of nodes contained in set $A$. $\tau^*$ and $\tau$ are the exponents of the cluster size distribution at $t_c$ obtained from an analytic formula~{\cite{cho}} and simulations, respectively. $\sigma^\prime$ is the exponent of the characteristic cluster size $s_A$. $\alpha$ and $\alpha^{\prime}$ are the exponents of the IET distribution in the time windows $[0,t_g]$ and $[t_a,t_g]$, respectively. $\delta_1$ and $\delta_2$ are the exponents of the jump size distribution for small and large $\Delta m$, respectively. The exponents $\eta$ and $\zeta$ are associated with growth of the largest cluster.} 
\label{table1}
\end{table*}

\end{document}